\documentclass[12pt]{article}
\usepackage{amsmath, amssymb}
\usepackage[dvips]{graphicx}
\usepackage{epsf}

\renewcommand{\thefootnote}{\fnsymbol{footnote}}

\begin{document}
\baselineskip=16pt

\begin{titlepage}

\def\thefootnote{\fnsymbol{footnote}}

\begin{center}

\hfill IPMU-11-0062\\
\hfill UT-11-10\\

\vskip .75in

{\Large \bf 
Studying Very Light Gravitino at the ILC
}

\vskip .75in

{\large Shigeki Matsumoto}$^{(1)}$
and
{\large Takeo Moroi}$^{(2,1)}$

\vskip 0.25in

$^{(1)}${\it 
IPMU, TODIAS, University of Tokyo, Kashiwa, 277-8583, Japan}

\vskip 0.07in

$^{(2)}${\it
Department of Physics, University of Tokyo, Tokyo 113-0033, Japan}

\end{center}
\vskip .5in

\begin{abstract}

A collider signal with a stable gravitino of ${\cal O}(10)$eV mass at the International Linear Collider (ILC) experiment is investigated. Such a light gravitino is generally predicted in the low-scale gauge mediation scenario of the supersymmetry breaking. We particularly focus on the case that the next lightest supersymmetric particle is stau, which eventually decays into a gravitino and a $\tau$-lepton. With such a small gravitino mass, the lifetime of the stau is $10^{-15}$--$10^{-11}$sec, and the produced stau decays before reaching the first layer of the inner detector of the ILC. It is shown, however, that the lifetime can be determined from the distribution of the impact parameter, which is obtained by observing charged tracks caused by decay products of the $\tau$-lepton. This measurement also enables us to estimate the mass of the gravitino and determine the scale of the supersymmetry breaking. Based on a simulation study, we found that the lifetime can be measured when it is longer than $\sim 10^{-14}$sec and the stau mass is about 100GeV.

\end{abstract}

\end{titlepage}

\renewcommand{\thepage}{\arabic{page}}
\setcounter{page}{1}
\renewcommand{\thefootnote}{\#\arabic{footnote}}
\setcounter{footnote}{0}

\section{Introduction}

The Large Hadron Collider experiment (LHC)~\cite{LHC} is now operating in order to clarify new physics beyond the standard model (SM). Though positive signals of the new physics are, unfortunately, not detected so far, those are expected to be found in near future, because the hierarchy problem of the SM strongly suggests that the new physics should be at the TeV scale or below. On the other hand, many new physics models have been theoretically proposed so far. Among those, the supersymmetric model~\cite{BookDrees} is the most attractive one, because it guarantees the stability of the Higgs mass against radiative corrections and gives a clue to solve the hierarchy problem of the SM. In addition, the supersymmetry (SUSY) plays a crucial role to realize the grand unification of known gauge interactions of the SM at a certain high energy scale.

Details of the supersymmetric model such as the mass spectrum of superpartners and their interactions are highly dependent of how the SUSY is broken, which is also particularly of importance to search supersymmetric signals at collider experiments. Several breaking mechanism are known~\cite{Intriligator:2007cp}. Among those, the gauge mediation scenario of the SUSY breaking~\cite{mGMSB} attracts attention, because it gives a solution to dangerous SUSY flavor problems which are generally caused by the introductions of superpartners of quarks and leptons. In the scenario, the breaking occurs at lower energy scale than those of other SUSY breaking models, so that the superpartner of the graviton, the gravitino, inevitably be the lightest supersymmetric particle (LSP). Here, we focus on the low-scale gauge mediation predicting the gravitino with ${\cal O}(10)$eV mass in this letter. Such a scenario with very light gravitino is well motivated because it is completely free from severe constraints coming from cosmology~\cite{Feng:2010ij} such as the Big-Bang Nucleosynthesis~\cite{BBN} and the formation of large scale structure of our universe~\cite{Viel:2005qj}.

In the study of the gauge mediation scenario of the supersymmetry breaking, the gravitino mass $m_{3/2}$ is a very important parameter because $m_{3/2}$ is directly related to the scale of the SUSY breaking, and also because the phenomenology and cosmology strongly depend on the gravitino mass. Thus, the experimental determination of the gravitino mass has a great impact on the study of the gauge mediation scenario. One way to determine the gravitino mass is to measure the decay width of superparticles into their superpartner and gravitino. In the gauge mediation scenario, the next lightest superparticle (NLSP) decays only into gravitino (and its superpartner), so the lifetime measurement of the NLSP gives a direct determination of the gravitino mass.

Collider signals of the low-scale gauge mediation scenario depend on what the NLSP is. Though there are many candidates for the NLSP, we focus on the stau NLSP in this letter, which is predicted in wide parameter region of the scenario. Once produced, the stau NLSP eventually decays into a $\tau$-lepton and a gravitino with the lifetime of $10^{-15}$--$10^{-11}$sec if the gravitino mass is of ${\cal O}(10)$eV, so that the decay length of the stau NLSP is much shorter than the typical size of the detectors of collider experiments. Thus, at the LHC experiment, for example, we expect the typical supersymmetric signal, namely, multi-jets associated with missing energy and $\tau$-leptons. Such a signal is, however, generally predicted in various SUSY breaking scenarios. Thus, even though the existence of SUSY is likely to be confirmed by the LHC experiment, the measurement of the stau lifetime looks challenging at the LHC.

The International Linear Collider (ILC)~\cite{ILC} is nothing but the experiment to pin down the new physics model under this kind of circumstance. The ideal environment with low background enables us to perform the precision study of the new physics. In this letter, we show that the measurement of the stau lifetime is possible by observing the distribution of the impact parameter, which is obtained by charged tracks caused by decay products of the $\tau$-lepton at the stau decay. In what follows, we consider how and how accurately the lifetime can be determined at the ILC based on a Monte-Carlo simulation study.

\section{Signal and Background Events}

The basic idea to measure the lifetime of the stau NLSP is the use of the simplest process producing the stau pair, $e^+e^- \to \tilde{\tau}^+\tilde{\tau}^-$, which eventually gives signal events involving two tau candidates with large impact parameters. In our analysis, we use only the events in which $\tau$-leptons decay hadronically in order to avoid the SM background from the process $e^+e^- \to W^+W^-$ followed by the decay of the $W$ boson into electron or muon (and neutrino).  In addition, we assume that the contamination of the misidentified $\tau$-jet-like object into the $\tau$-jet sample is negligible.

\begin{figure}[t]
\begin{center}
\includegraphics[scale=0.45]{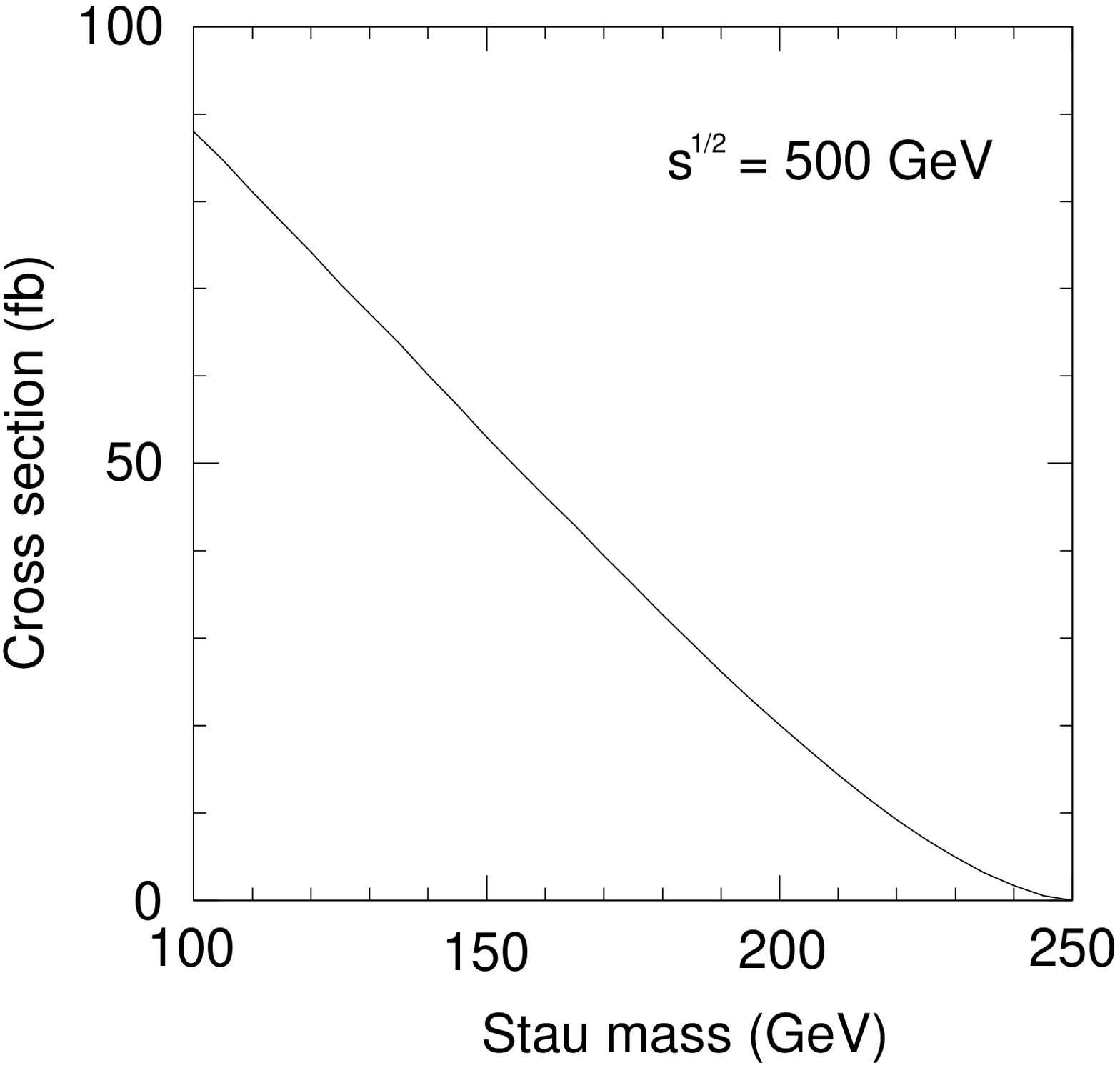}
~~~~
\includegraphics[scale=0.45]{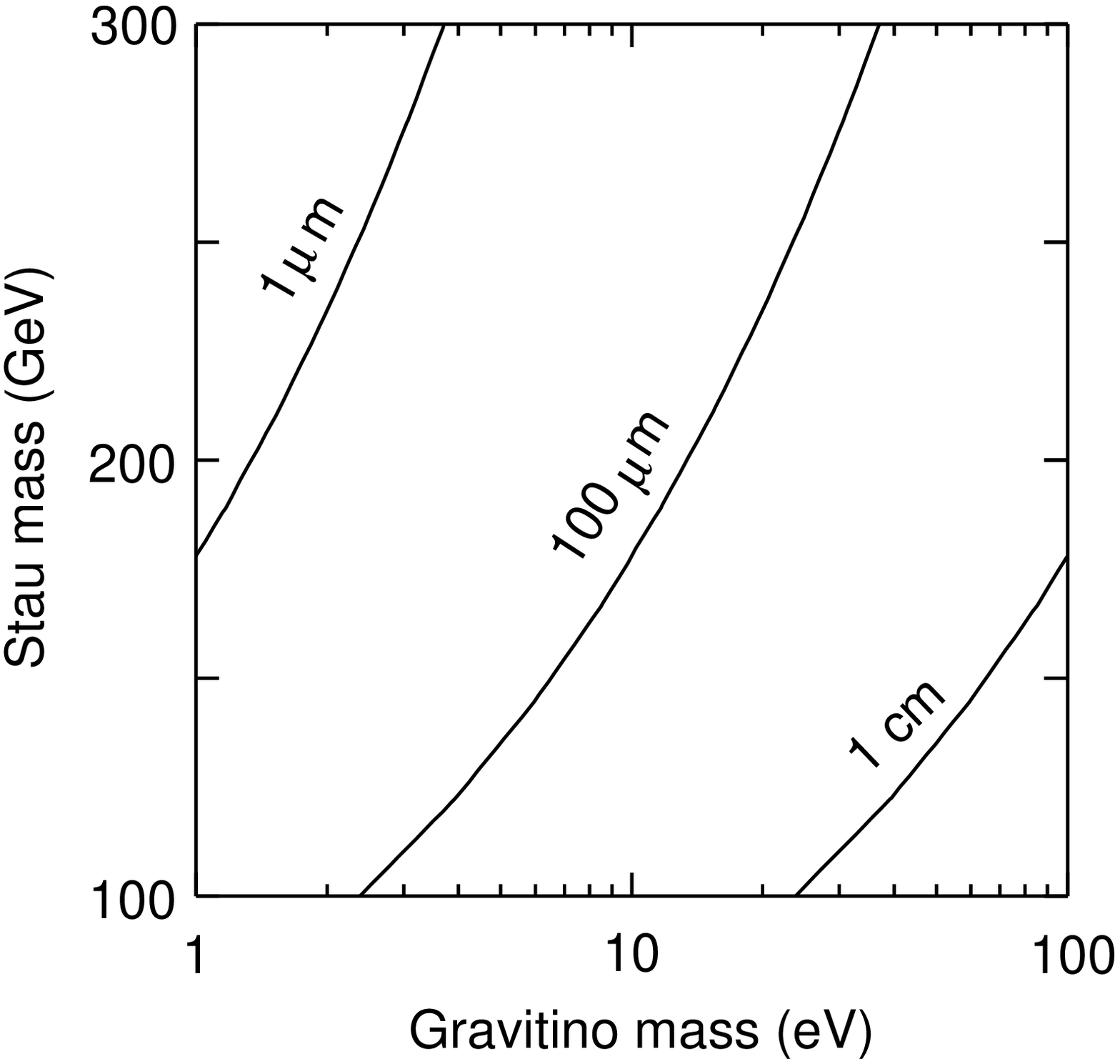}
\caption{\small Cross section of the signal process, $e^+e^- \to \tilde{\tau}^+\tilde{\tau}^-$, as a function of $m_{\tilde{\tau}}$ (left panel) and the contour plot of the decay length, $c\tau_{\tilde{\tau}}$, on the ($m_{3/2}$, $m_{\tilde{\tau}}$)-plane (right panel). The lifetime of the stau NLSP is given by $\tau_{\tilde{\tau}} \simeq 3.34 \times 10^{-13} (c\tau_{\tilde{\tau}}/100\mu{\rm m})$ sec.}
\label{fig:signal}
\end{center}
\end{figure}

The cross section of the signal process, $e^+e^- \to \tilde{\tau}^+ \tilde{\tau}^-$, is depicted in the left panel of Fig.~\ref{fig:signal} as a function of the stau mass $m_{\tilde{\tau}}$. It can be seen that enough number of stau pairs are produced at the ILC. Stau NLSP decays into a $\tau$-lepton and a gravitino once it is produced, and its lifetime is determined by the formula,
\begin{eqnarray}
\tau_{\tilde{\tau}}
=
48 \pi M_{\rm pl}^2~m_{3/2}^2 / m_{\tilde{\tau}}^5
\simeq
5.9 \times 10^{-12}{\rm sec}
\times
\left( \frac{m_{3/2}}{10{\rm eV}} \right)^2
\left( \frac{m_{\tilde{\tau}}}{100{\rm GeV}} \right)^{-5},
\end{eqnarray}
where $M_{\rm pl} \simeq 2.44 \times 10^{18}$GeV is the reduced Planck mass, while $m_{3/2}$ is the gravitino mass. When the gravitino mass is of the order of 10eV and the stau is lighter than 200GeV, the decay length of the stau turns out to be longer than that of $\tau$-lepton ($c\tau_\tau \simeq 87.11\mu$m~\cite{Nakamura:2010zzi}) as shown in the right panel of Fig.~\ref{fig:signal}.

Since the stau NLSP decays before reaching the first layer of the inner detector, which is now designed to locate at 16mm away from the beam pipe~\cite{:2010zzd}, the charged track of the stau cannot be seen. Instead, we focus on the distribution of the impact parameter, which is the shortest distance to the track from the interaction point, of the charged tracks arising from the stau decay.\footnote{In our study, we neglect the effect of the magnetic field inside the detector, so that all the tracks of charged particles are approximated to be straight. Even if the charged tracks are not exactly straight in the actual experimental circumstance, the lifetime of the stau NLSP can be measured at the ILC as far as the profile of the magnetic field inside the detector is well understood. More complete analysis based on a full simulation study will be given elsewhere~\cite{FMMS}.} Using the distribution of the impact parameter, we are able to measure the lifetime of the stau NLSP, which leads to the determination of the gravitino mass and therefore the scale of the SUSY breaking.

If the decay length of the stau NLSP is much longer than that of the $\tau$-lepton, SM backgrounds are easily eliminated by requiring $\tau$-jets with large impact parameters. The SM backgrounds become, on the contrary, important when the decay length of the stau NLSP is the same order or shorter than that of the $\tau$-lepton.

We expect large amount of SM background events including two $\tau$-jets. In the study of the impact-parameter distribution of the $\tau$-jets from the $\tilde{\tau}$ decay, significant sources of SM backgrounds are expected to be coming from the processes,
\begin{itemize}
\item[(a)] [$\tau\tau$-BG] $e^+e^- \to \tau^+\tau^-$ ($+$ ISR),
\item[(b)] [$WW$-BG] $e^+e^- \to W^+W^- \to \tau^+\tau^-\nu\bar{\nu}$ ($+$ ISR),
\item[(c)] [$ZZ$-BG] $e^+e^- \to Z^0Z^0 \to \tau^+\tau^-\nu\bar{\nu}$ ($+$ ISR),
\end{itemize}
where ISR denotes the initial state radiation. Cross sections of these backgrounds are 1144fb, 99fb, and 6fb for $\tau\tau$-BG, $WW$-BG, and $ZZ$-BG, respectively, when the center of mass energy at the collision is $\sqrt{s}=500$GeV. Notice that the cross sections for the $WW$- and $ZZ$-BG include relevant leptonic branching ratios of weak bosons.

It should be noted that $\tau$-pair may be also produced by the two photon event $\gamma\gamma\rightarrow \tau^+\tau^-$, where photons are emitted from the initial-state $e^\pm$. This type of background is, however, eliminated if an initial-state $e^\pm$ is detected; with the currently expected performance of the ILC detector, the forward region is covered up to $|\eta| \leq 5.22$~\cite{Bambade:2006qc}, which will help to reduce the background. In addition, the two photon event is suppressed as we impose a lower bound on the energy of $\tau^+\tau^-$ system. In our analysis, we impose a lower bound on the total visible energy in the event, and neglect the background from the two photon event. Following Ref.~\cite{Suehara:2009nj}, we require that $E_{\rm vis}$, which is defined by the total energy of charged particles in the event, should be larger than 70GeV. (See the Cut 2 in the next section.)

Now we discuss how we can eliminate $\tau\tau$-BG, $WW$-BG, and $ZZ$-BG. One important observation in reducing backgrounds is that, in the pair production processes (a) -- (c), $\tau^\pm$, $W^\pm$, and $Z^0$ are likely to be produced with high rapidity. Then, in the $WW$-BG and $ZZ$-BG, $\tau^\pm$ are produced by the decay of weak bosons. Even in such cases, the information about the direction of weak bosons are inherited in the direction of decay products, i.e., $\tau^\pm$. Thus, concentrating on events with $\tau$-clusters with low rapidity, standard-model backgrounds are significantly reduced. (See the Cut 3.) Because the cross section for the process $e^+e^-\rightarrow \tau^+\tau^-$ ($+$ ISR) is significantly larger than the signal cross section, the rapidity cut on $\tau$-clusters is not enough to reduce $\tau\tau$-BG. Large amount of remaining $\tau\tau$-BG can be eliminated by using the fact that, in the $\tau\tau$-BG, two $\tau$-leptons are (almost) back-to-back. If significant amount of momentum is carried away by ISR photons, $\tau^+$ and $\tau^-$ are not back-to-back in general. However, even in such events, the ISR photon is likely to be (almost) parallel to the initial beam direction. Thus, we reject events in which two $\tau$-clusters projected on the $x$-$y$ plane are back-to-back. (See the Cut 4.) We found that such cut is also useful to reduce the $WW$-BG. In addition, if an ISR photon is emitted with large transverse momentum, momenta of two $\tau$-leptons and that of the photon should be on one plane in the $\tau\tau$-BG event. This fact can be also used to reduce the $\tau\tau$-BG. (See the Cut 5.)

\section{Simulation Results}

In this section, we present several results of our simulation study. We have generated both the signal and background events using the HERWIG package~\cite{HERWIG}. This package enables us to generate the events not only with momentum information but also those of decay points. In our study, we fix the stau mass to be $m_{\tilde{\tau}} =$ 120GeV. We neglect the uncertainty in the stau mass because the stau mass is expected to be determined with the accuracy less than 1\% by measuring the energy distribution of reconstructed $\tau$-leptons from the decay of the stau NLSP~\cite{StauMass} at the signal process.

For each events, we identify decay products of $\tau$-leptons. The efficiency of the $\tau$-jet tagging is assumed to be 100\%. For $\tau$-jet labeled by $I$, which is defined by hadronic decay products of $I$-th $\tau$-lepton, we define the momentum of the $\tau$-jet (${\bf P}_I$) to be
\begin{eqnarray}
{\bf P}_I = \sum_{i = \pi^\pm, K^\pm, \cdots} {\bf p}_I^{(i)},
\end{eqnarray}
where ${\bf p}_I^{(i)}$ is the three momentum of the particle $i$ in the $I$-th $\tau$-jet, and the summation is over all the long-lived charged mesons in $I$-th $\tau$-jet. ${\bf P}_I$ is parameterized as
\begin{eqnarray}
{\bf P}_I
=
(P_I \sin\theta_I \cos\phi_I, P_I \sin\theta_I \sin\phi_I, P_I \cos\theta_I).
\label{tau jet momentum}
\end{eqnarray}

\begin{table}
\centering
\begin{tabular}{l|rrrr} 
& Signal & $\tau\tau$-BG & $WW$-BG & $ZZ$-BG \\
\hline
Cut 1+2 & $2630$ & $45172$ & $3911$ & $247$ \\
Cut 1+2+3 & $2197$ & $1798$ & $911$ & $79$ \\
Cut 1+2+3+4 & $1316$ & $525$ & $362$ & $26$ \\
Cut 1+2+3+4+5 & $1307$ & $116$ & $353$ & $26$ \\
\hline
\end{tabular}
\caption{\small The cut statistics for background suppression. Here, the number of events is normalized to the integrated luminosity ${\cal L} = $100fb$^{-1}$ with the center of mass energy $\sqrt{s} =$ 500GeV. Mass of the stau NLSP is set to be 120GeV.  In our analysis, the efficiency of the signal event is insensitive to $\tau_{\tilde{\tau}}$ as far as the decay length of the stau is much shorter than the detector size, which is the case in the present setup.}
\label{table:cutstat}
\end{table}

Then, based on the discussion given in the previous section, we impose the following kinematical cuts on the generated events to reduce the SM backgrounds:
\begin{itemize}
\item[1.] The number of $\tau$-jet should be equal to 2.
\item[2.] $E_{\rm vis} \geq$ 70GeV.
\item[3.] $|\cos\theta_1| \leq 0.85$ and $|\cos\theta_2| \leq 0.85$.
\item[4.] $-0.95\leq\cos(\phi_1-\phi_2) \leq 0.25$.
\item[5.] If there exist isolated photons with transverse momentum larger than 20GeV, we require $|\hat{\bf P}_1 \times \hat{\bf P}_2 \times \hat{\bf p}_{\gamma}^{\rm (1)}|\geq 0.1$. (Here, ${\bf p}_{\gamma}^{\rm (1)}$ is the momentum of highest $p_T$ isolated photon, and $\hat{\bf a}$ is the unit vector parallel to ${\bf a}$.)
\end{itemize}
In order to see how well the SM backgrounds are eliminated, and hence to see how well we can determine $\tau_{\tilde{\tau}}$, by imposing the kinematical cuts, we show the cut-table (the cut-statistics for background reduction) in Table\ \ref{table:cutstat}. It can be seen that the number of background events can be made small enough by imposing the kinematical cuts.

With the use of the generated events which pass the above kinematical cuts 1--5, we calculate the impact parameter of each $\tau$-jet, which is given by
\begin{eqnarray}
b_I \equiv
\left|
{\bf x}_I - \frac{{\bf x}_I \cdot {\bf P}_I}{|{\bf P}_I|^2}{\bf P}_I
\right|,
\label{Impact parameter}
\end{eqnarray}
where ${\bf x}_I$ is the decay point of the $I$-th $\tau$-lepton. Note that the summation over the index $I$ should not be taken here. We expect that, in the realistic circumstance of the ILC experiment, the distribution of the impact parameter $b_I$ is obtained by measuring the shortest distance to the $\tau$-jet track from the interaction point.

\begin{figure}[t]
\begin{center}
\includegraphics[scale=0.45]{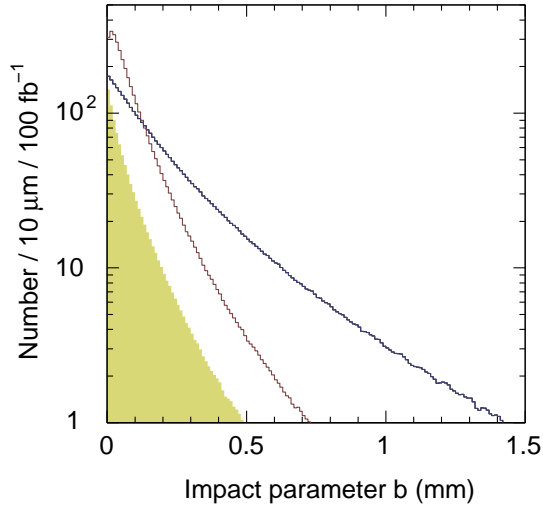}
\caption{\small Distribution of the impact parameter for the signal and background events. The solid histograms are distributions of signals for $\tau_{\tilde{\tau}} = 10^{-13}$ (red) and $10^{-12}$sec (blue), while the background is given by the shaded one. The stau mass is fixed to be 120 GeV.}
\label{fig:bdist}
\end{center}
\end{figure}

The resultant distributions of the impact parameter for the signal and background events are shown in Fig.~\ref{fig:bdist}. We adopt the error of the impact-parameter measurement of 5$\mu$m here and hereafter according to Ref.~\cite{:2010zzd}. It can be seen that the typical size of the impact parameter of the background events are $\sim$ 100$\mu$m, which is expected from the order of the decay length of the $\tau$-lepton. On the other hand, the typical size of the impact parameter for the signal event depends strongly on the lifetime $\tilde{\tau}$. This fact indicates that the lifetime of the stau NLSP is determined accurately once the distribution of the impact parameter is obtained accurately.

\begin{figure}[t]
\begin{center}
\includegraphics[scale=0.45]{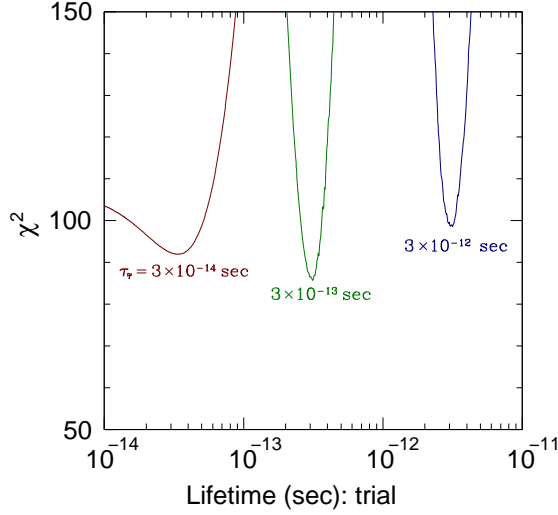}
\caption{\small The $\chi^2$-variable of the ${\cal L} =$ 100fb$^{-1}$ data for the three cases of the stau lifetime; $3 \times 10^{-14}$sec, $3 \times 10^{-13}$sec, and $3 \times 10^{-12}$sec. Stau mass is fixed to be 120 GeV.}
\label{fig:chi2}
\end{center}
\end{figure}

Now, we quantitatively estimate how accurately the lifetime can be determined by using the distribution of the impact parameter. The number of event in $i$-th bin of the distribution is denoted as $N_i(\tau_{\tilde{\tau}})$; in our analysis, we have used the bins for 0 $\leq b \leq$ 2mm with the bin-width of 20$\mu$m, so that the number of bins is $100$. In order to see how well the lifetime can be constrained, we also calculate the theoretical expectation of the number of events in the $i$-th bin $N_i^{\rm(th)}(\tau_{\tilde{\tau}})$ as a function of $\tau_{\tilde{\tau}}$. Here, $N_i^{\rm (th)}(\tau_{\tilde{\tau}})$ has been obtained by using the events generated with very high statistics. Then, for a given underlying value of the stau lifetime $\tau_{\tilde{\tau}}$, we calculate the $\chi^2$-variable as a function of the postulated value of the lifetime (denoted as $\tau_{\tilde{\tau}}^{\rm (trial)}$):
\begin{eqnarray}
\chi^2(\tau_{\tilde{\tau}}; \tau_{\tilde{\tau}}^{\rm (trial)})
\equiv
\frac{1}{N_i^{\rm (th)}(\tau_{\tilde{\tau}}^{\rm (trial)})}
\sum_i
\left(
N_i(\tau_{\tilde{\tau}}) 
-
N_i^{\rm (th)}(\tau_{\tilde{\tau}}^{\rm (trial)})
\right)^2,
\label{chi^2}
\end{eqnarray}
where $i$ is summed over all the bins for the distribution of the impact-parameter.

Taking ${\cal L} =$ 100fb$^{-1}$, in Fig.~\ref{fig:chi2}, we have shown the $\chi^2$ variable for three cases of the underlying stau lifetime; $3 \times 10^{-14}$sec (red line), $3 \times 10^{-13}$sec (green line), and $3 \times 10^{-12}$sec (blue line). The minimum of the $\chi^2$ value divided by the degree of the freedom $(100 - 1 = 99)$ is 0.93, 0.87, and 0.99, respectively. On the other hand, since $\Delta\chi^2 (\equiv \chi^2 - \chi^2_{\rm min}) =$ 1 corresponds to the observed lifetime with 68\% C.L., the figure is showing that the lifetime of the stau NLSP can be measured to be $(3.36^{+0.62}_{-0.62}) \times 10^{-14}$sec, $(3.13^{+0.06}_{-0.15}) \times 10^{-13}$sec, and $(3.16^{+0.07}_{-0.17}) \times 10^{-12}$sec, respectively.

\begin{figure}[t]
\begin{center}
\includegraphics[scale=0.45]{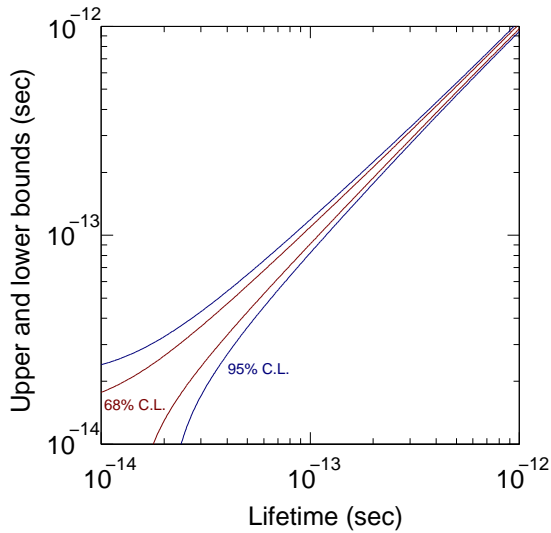}
\caption{\small Measurement accuracies of the lifetime of the stau NLSP. Upper and lower bound on the lifetime at 68\% (red lines) and 95\% (blue lines) confidence levels are shown as a function of the lifetime. The stau mass is fixed to be 120 GeV.}
\label{fig:dchi2}
\end{center}
\end{figure}

We also consider the measurement accuracy of the lifetime of the stau NLSP. For this purpose, we estimate the $\Delta\chi^2$ by replacing $N_i(\tau_{\tilde{\tau}}) \rightarrow N_i^{\rm(th)}(\tau_{\tilde{\tau}})$ in Eq.\ \eqref{chi^2}. Based on this quantity, we derive 68\% and 95\% confidence levels of upper and lower bounds on the lifetime. The result is shown in Fig.~\ref{fig:dchi2}. It can be seen that the nonzero lifetime of the stau NLSP can be confirmed when the lifetime is longer than $\sim 10^{-14}$sec. The measurement accuracy becomes better when the lifetime becomes longer. For instance, the lifetime can be measured with the accuracy less than 5\% when $\tau_{\tilde{\tau}} > 10^{-12}$sec.

\section{Summary and discussions}

We have studied a collider signal with a stable gravitino of ${\cal O}(10)$eV mass at the ILC. In our analysis, we particularly focus on the case that the NLSP is stau, which eventually decays into a gravitino and a $\tau$-lepton with the lifetime of $10^{-15}-10^{-11}$sec. It has been found that the signal of the stau decay can be detected by measuring the distribution of the impact parameter, which is obtained by observing charged tracks caused by decay products of the $\tau$-lepton. We have also found that the lifetime can be measured with enough accuracy when it is longer than $\sim 10^{-14}$sec and the stau mass is about 100 GeV. This fact guarantees the capability of the ILC to test the low-scale gauge mediation scenario of the SUSY breaking.

In our study, we only consider the distribution of the impact parameter at the ILC with the center of mass energy 500 GeV to measure the lifetime of the stau NLSP. It may be possible to increase the significance to detect the signal by choosing appropriate center of mass energy according to the stau mass. In addition, the use of the incident polarized electron/positron beam will help us to reduce the background significantly. Furthermore, observing not only the impact parameter but also the direct information about the decay point of the $\tau$-lepton using 3-prong hadronically decay may enable us to measure the lifetime smaller than $\sim 10^{-14}$sec. These kinds of analysis require us to go to a full simulation of the detector performance and it is beyond the scope of this letter though those will be given in near future~\cite{FMMS}.

Finally, we comment on the stau NLSP at the LHC. In the model of a long-lived stau, we also expect sizable number of SUSY events which contain $\tau$-jets with large impact parameters even at the LHC experiment. Although the full reconstruction of the SUSY event is challenging at the LHC, the determination of the lifetime of the stau NLSP may be possible. More interestingly, at the early stage of the LHC experiment, the discovery reach of the SUSY may be extended by looking for $\tau$-jets with large impact parameter. These subjects will be discussed elsewhere~\cite{StauLHC}.

\section*{Acknowledgments}

We are deeply grateful to K. Fujii, T. Suehara and K. Yonekura for useful discussions and comments. This work is supported by Grant-in-Aid for Scientific research from the Ministry of Education, Science, Sports, and Culture (MEXT), Japan, Nos.\ 21740174 \& 22244031 (S.M.), No.\ 22540263 (T.M.), and No.\ 22244021 (S.M. and T.M.), and also by World Premier International Research Center Initiative (WPI Initiative), MEXT, Japan.

\end{document}